\documentclass[preprint]{revtex4}  
\usepackage{amsfonts}
\usepackage{amsmath}
\usepackage{cancel}
\usepackage{graphicx}
\usepackage{mathcomp}
\usepackage{xcolor}
\usepackage{ulem}
\usepackage{enumitem}

\newif\iffigures
\figurestrue

\renewcommand{\vec}{\boldsymbol}

\renewcommand{\vec}{\boldsymbol}

\def\undertilde#1{\mathord{\vtop{\ialign{##\crcr
$\hfil\displaystyle{#1}\hfil$\crcr\noalign{\kern1.5pt\nointerlineskip}
$\hfil\widetilde{}\hfil$\crcr\noalign{\kern1.5pt}}}}}

\begin{document}

\title{Kinetic Simulation of the Ideal Multipole Resonance Probe}

\author{Junbo Gong$^{1}$, Michael Friedrichs$^{2}$, Jens Oberrath$^{2}$, and Ralf Peter Brinkmann$^{1}$}
\affiliation{$^{1}$ Ruhr University Bochum\\Department for Electrical Engineering and Information Technology
Institute of Theoretical Electrical Engineering\\ 
D-44801 Bochum, Germany\\
$^{2}$ South Westphalia University of Applied Science\\
Department of Electrical Power Engineering\\
Modeling and Simulation\\
D-59494 Soest, Germany}

\date{\today}

\renewcommand{\abstractname}{}

\begin{abstract}
\vspace*{0.5cm}
\textcolor[rgb]{0.1,0.1,0.1}{\textit{Active plasma resonance spectroscopy (APRS)} is a process-compatible plasma diagnostic method which utilizes the natural ability of plasmas to resonate on or near the electron plasma frequency. The \textit{Multipole Resonance Probe (MRP)} is a particular design of \textit{APRS} that has a high degree of geometric and electric symmetry. The principle of the \textit{MRP} can be described on the basis of an idealized geometry that is particularly suited for theoretical investigations. In a pressure regime of a few Pa or lower, kinetic effects become important, which can not be predicted by the Drude model. Therefore, in this paper a dynamic model of the interaction of the idealized MRP with a plasma is established. The proposed scheme reveals the kinetic behavior of the plasma that is able to explain the influence of kinetic effects on the resonance structure. Similar to \textit{particle-in-cell}, the spectral kinetic method iteratively determines the electric field at each particle position, however, without employing any numerical grids. The optimized analytical model ensures the high efficiency of the simulation. Eventually, the presented work is expected to cover the limitation of the Drude model, especially for the determination of the pure collisionless damping caused by kinetic effects. A formula to determine the electron temperature from the half-width $\Delta\omega$ is proposed.
}
 
\end{abstract}

\maketitle

\pagebreak

%
%
\section{Introduction}
\begin{figure}[!htb]	
  \includegraphics[width=0.9\textwidth]{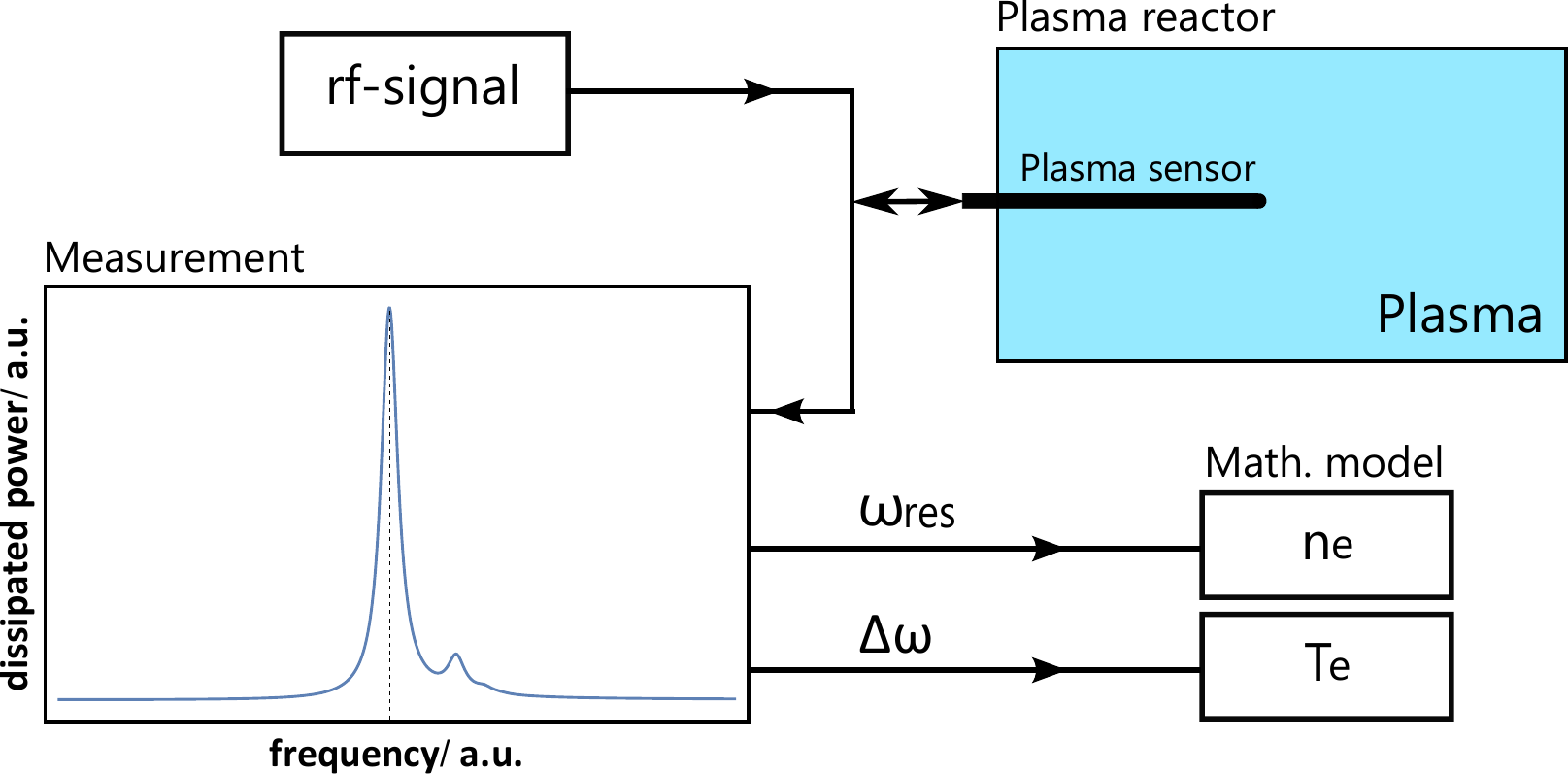} 
	\caption{Schematic depiction of APRS: an rf-signal is coupled into a plasma and the spectral response is recorded. From the resonance frequency and $\Delta \omega$, the corresponding electron density and electron temperature can be determined by using a mathematical model.}
	\label{fig:APRS}
  \end{figure}

An essential part of the plasma technology is the supervision and control of plasmas in industrial applications. One of the industry-compatible approaches to plasma diagnostics is the so-called active plasma resonance spectroscopy (APRS) \cite{Tonk and Langmuir}. As shown in Fig.~\ref{fig:APRS}, the idea of APRS is to couple an RF signal into the plasma via a probe and to measure the response of the signal in a certain frequency range. The absorption peaks are identified where electrons resonate. Then a specific mathematical model is used to determine some important plasma parameters such as electron density $n_{\rm{e}}$ or electron temperature $T_{\rm{e}}$. This concept was initially discussed and investigated back in 1929, and it has been applied and analyzed in many different designs since then \cite{TakayamaMiyazaki1960,Levitskii1963,Buckley1966,Stenzel1976,PiejakGodyak2004,Dine2005,Scharwitz2009,XuSugai2009,LiLiu2010,WangLiu2011,LinagSugai2011,SchulzStyrnollAwakowiczRolfes2015,Nakamura1999,Nakamura2019}.

\begin{figure}[!htb]	
  \includegraphics[width=0.9\textwidth]{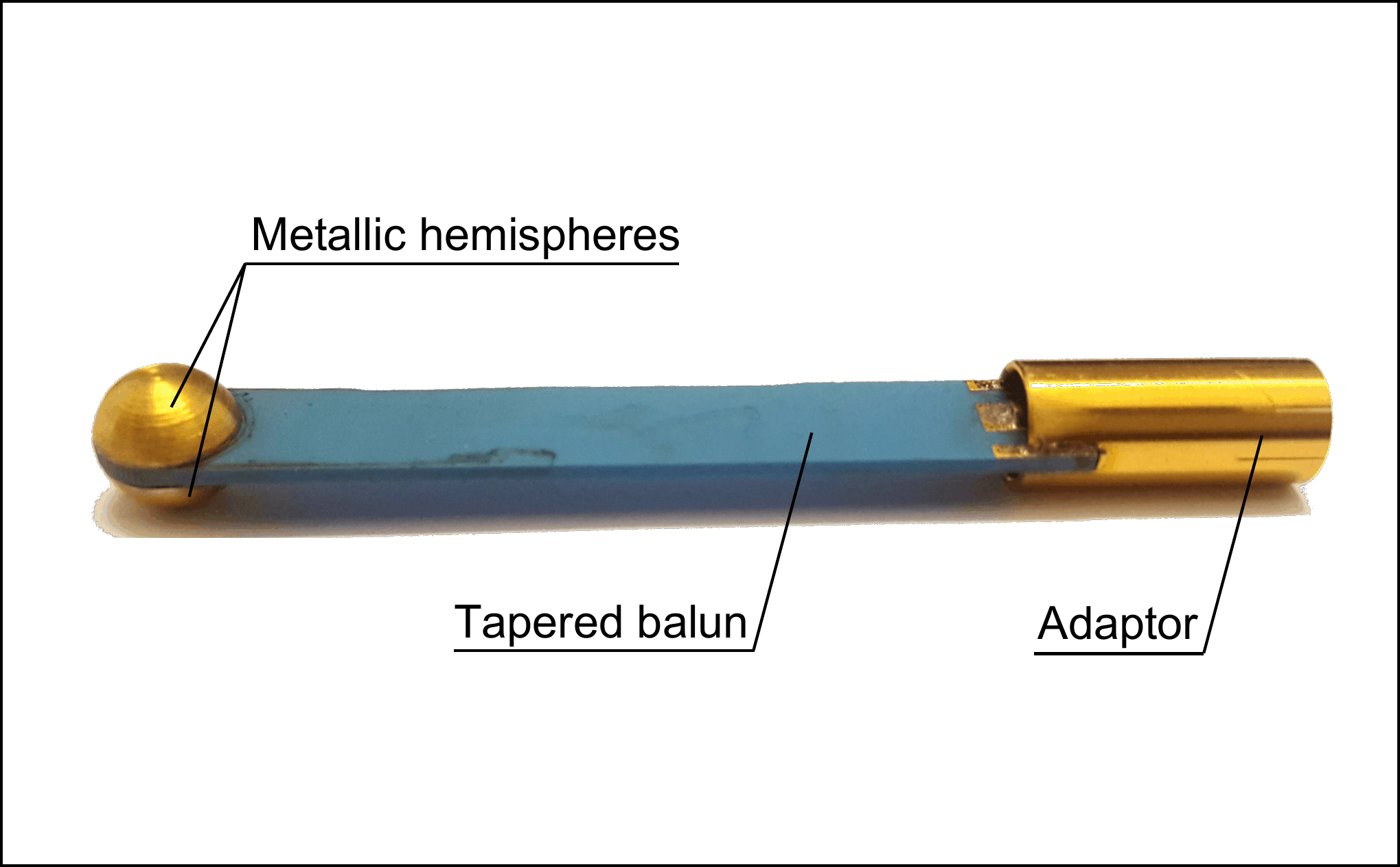} 
	\caption{Prototype of the MRP: The probe consists of two metallic hemispheres with a total diameter 8$\rm{mm}$. It is symmetrically driven via a tapered balun transformer and it can be covered in a cylindrical quartz tube.}
	\label{fig:MRP_SCHEMATIC}
  \end{figure}

The Multipole Resonance Probe (MRP) is one of the realizations of the APRS. The prototype of the MRP is shown in Fig.~\ref{fig:MRP_SCHEMATIC}: it consists of a spherical probe head and a holder. The probe head is comprised of two dielectrically shielded, conducting metallic hemispheres which constitute the electrically symmetric electrodes. The hemispheres are fixed to a holder that contains the RF-supply. The setup of the MRP holds two important features: its geometry is approximately symmetric, and its electrical behavior is symmetric with respect to the mapping. 

To understand the behavior of the surrounding plasma and the influence of the probe, the Drude model of the MRP has been studied, where the plasma is treated as a frequency-dependent dielectric material. In \cite{Lapke2011,LapkeMussenbrockBrinkmann2008}, the analytical evaluation of the resonance spectrum for the idealized MRP has been derived, where the holder is neglected. In \cite{SchulzStyrnollOberrath2013}, the numerical simulation of the MRP with complete geometry is investigated within 3D-electromagnetic field simulations using CST Microwave Studio. The influence of the holder is included to provide accurate results. However, both approaches demonstrate very similar resonance behavior of the plasma: the identical dipole resonance peak is dominant. It indicates the feasibility of the idealized model. In \cite{SchulzRolfes2014}, the simulated magnitude of the electric field shows that most of the interactions in the plasma appear near the tip of the probe. Hence, the hemisphere on the opposite side of the holder can be seen as an ideal model and the other half of the sphere is assumed symmetric. Then the ideal MRP (IMRP) is formed with a geometric symmetry: two ideal hemisphere electrodes are covered by a dielectric. 

The theoretical and numerical results are substantially confirmed through the simulation and experimental work \cite{JensBrinkmann_Eigen2014,StyrnollAwa2013,StyrnollLapkeAwa2014,LapkeOberrathMussenbrockBrinkmann2013}. Nevertheless, there are deviations: the fluid approach has a poor prediction in a pressure regime of a few Pa and lower. While the absolute position of the resonances was quantitatively recorded, the half-width $\Delta \omega$ of dependent damping cannot be reproduced. In the Drude model, the energy loss strongly depends on the electron-neutral collisions. However, in reality collisionless loss exists because the electrons can be deflected out of the influenced domain by the electric field, which is missing in the Drude model. To overcome the limitation, a kinetic investigation is required. In \cite{Jens_APRS,Jens_PEP,Jens_IP,Jens_MRP}, a general kinetic model of the probe-plasma system is discussed, and the functional analytic solutions of its specific geometries are determined. Unfortunately, it remains challenging for the collisionless case, which appears to be greatly relevant to reveal the pure kinetic effects. Alternatively, solving the integration of the Vlasov equation is a possible approach. However, it is cumbersome in terms of calculations. Therefore, a more straightforward particle model is necessary.

\begin{figure}[!htb]
\centering
\includegraphics[width=0.9\textwidth]{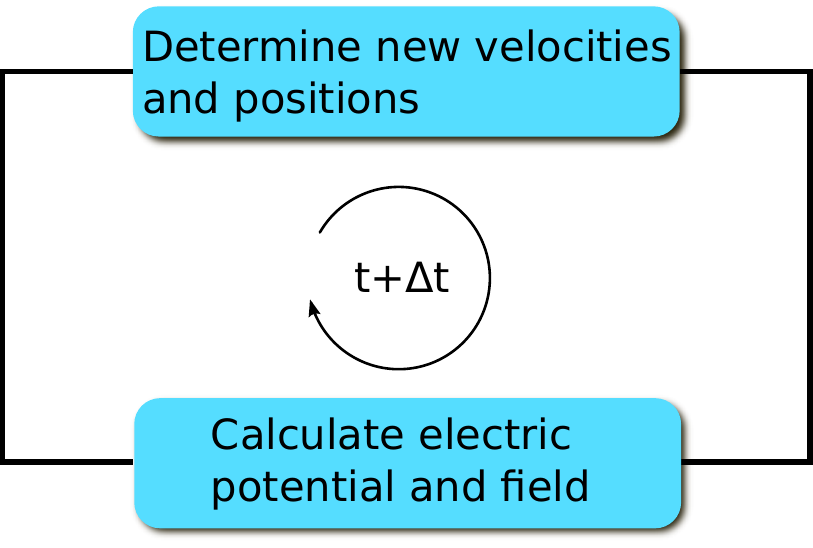}
\caption{Scheme of the spectral kinetic simulation. The particles move individually according to the results of the dynamic equations in each iteration. In this paper, we consider a collisionless case in our investigation.}
\label{fig:Set-up}
\end{figure}

In fact, particle-based models are becoming the focus of research due to their excellent scalability with the development of computer science for many years~\cite{Verlet1967,BirdsallLangdon1985}, such as the particle-in-cell (PIC) method, which was first proposed for modeling compressed fluids~\cite{PIC}. In~\cite{Hellberg1968, Calder1985}, the simulation of diagnostic devices was presented using a method similar to PIC. Although PIC is widely used for plasma modeling, full PIC codes are computationally expensive and may cause statistical noise. Thus, the spectral kinetic method is developed to efficiently capture the kinetic effects. Fig.~\ref{fig:Set-up} shows the simulation scheme, which consists of two modules: particle pusher and field solver. The dynamics of the self-consistent system are described in the Hamiltonian formalism, and the Poisson problem can be solved with a Green's function. It determines the electric field at each particle position iteratively without employing any numerical grid cells. The spherically symmetric geometry of IMRP simplifies the calculations for solving the kinetic model. 

In reality, the plasma systems are extremely large concerning the number of particles. To enhance the efficiency of the simulation, so-called ``super-particles'' can be applied, where a certain number of real particles are presented by a single super-particle. However, it affects the characterization of the plasma, such as the exaggerated coulomb force between super-particles in a short range. In PIC, the information of the super-particles in the same cells is communicated via interpolation, which implies the vanishing interactions between particles at short distances in the same numerical cell. Since these cells are not defined in the spectral kinetic scheme, a certain truncation of infinite series expansions of the Green's function has to be made, which leads to an optimized mathematical model.

The spectral response of the system is expected to demonstrate the influence of kinetic effects. However, the calculations in the frequency domain are time-consuming regarding the convergence of a sequence of periodic functions. In the spectral kinetic scheme, an efficient approach is proposed: an impulse signal is provided as the input of the system, then the charge on the electrodes in the time domain are recorded as the output. Applying a Fourier transformation, the impulse response can be expressed in the frequency domain. Eventually, by comparing with the Drude model, the effective collision frequency is obtained in the kinetic model, which can be used to determine the electron temperature.

The paper is structured as follows: Chapter 2 gives a general description of the plasma-IMRP system. In chapter 3, the spectral kinetic scheme is introduced analytically in detail, and it is simplified due to the symmetry of the IMRP. In chapter 4, the implementation is presented regarding the discussion of the initial condition, boundary condition, and simulation results. Finally, the paper concludes with chapter 5.

\pagebreak

\section{Idealized MRP model}
\begin{figure}[!htb]
\centering
\includegraphics[width=0.9 \textwidth]{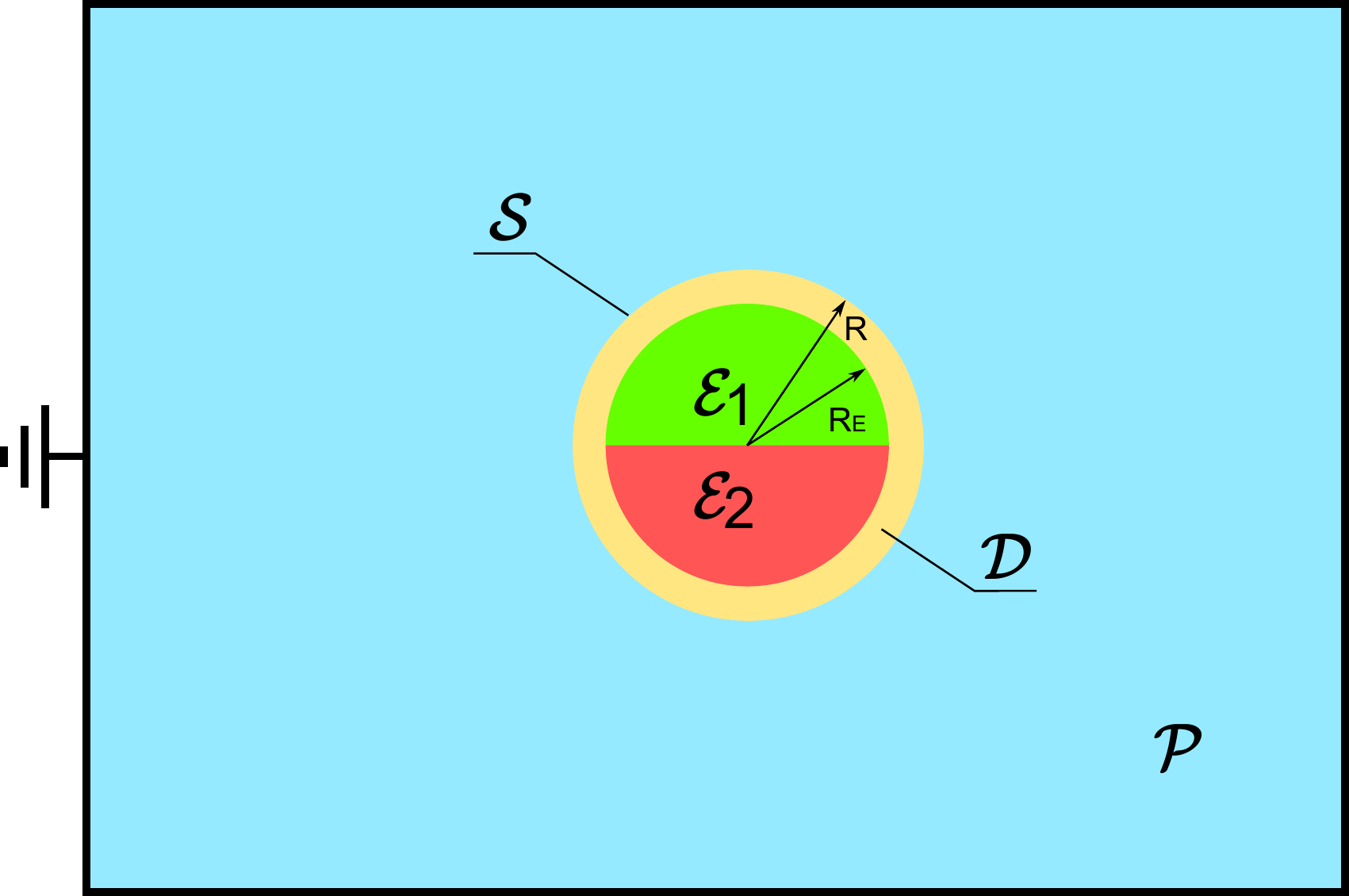} 
\caption{Schematic of the ideal axially symmetrical model of the MRP inside a plasma: the holder is neglected which is suitable for the theoretical investigation.}
\label{fig:Ideal}
\end{figure}

In Fig.~\ref{fig:Ideal}, the IMRP is illustrated: The chamber is given as a spatially bounded domain~$\mathcal{V}$, and IMRP, which consists of two ideal electrodes, is immersed in the plasma volume~$\mathcal{P}$: The ideal electrodes $\mathcal{E}_{1}$ and $\mathcal{E}_{2}$ with infinite conductivity are dielectrically covered in~$\mathcal{D}$. The voltages $u_1$ and $u_2$ can be applied to the electrodes which excite the system. The radius of the IMRP is $R$, and the thickness of the dielectric is $Rd$, which gives the radius of the electrodes $R_{\rm{E}}=R-Rd$. The plasma around the probe is treated as an ensemble of $N$ classical free point charges. The charge density $\rho(\vec{r})$ is given including the constant surface charge density $\sigma_{\mathcal{S}}$, which indicates the total surface charge $Q_{\mathcal{S}}$ is homogeneously distributed on $\mathcal{S}$. The electrostatic approximation $\vec{E}=-\nabla\Phi$ is adopted in~$\mathcal{V}$. The skin effect is negligible because the length scale is small compared to the skin depth, and no electromagnetic waves are emitted due to the fact that the frequency of the applied signal~$\omega_{\rm{rf}}$ is smaller than the electron plasma frequency $\omega_{\rm{pe}}$ \cite{MussenbrockHemkeKlick2008}. 

In IMRP-plasma system, the electric potential can be calculated in Poisson's equation, where index $n$ (or $n^{\prime})=1,2$ refers the corresponding electrode
\begin{align}\label{eqn:Poisson}
      -\nabla \!\cdot\! \left(\varepsilon_{\rm r}(\vec{r} ) \nabla \Phi(\vec{r})\right) &= \rho(\vec{r}), \\
	     \Phi(\vec{r}) &= \begin{cases}
	         0, \quad |\vec{r}|\to\infty\\
	         u_n, \quad  \vec{r} \in \mathcal{E}_n \nonumber\\
	    \end{cases}.
\end{align}
A suitable tool for the formal description of these relations is the Green's function $G(\vec{r},\vec{r}^{\,\prime})$, which gives 
\begin{alignat}{2}\label{eqn:Green}
    - \nabla  \cdot\! \left( \varepsilon_{\rm r}(\vec{r} ) \nabla  G(\vec{r},\vec{r}^{\,\prime})\right)
	&= \delta^{(3)}(\vec{r}\,-\vec{r}^{\,\prime}),\\
	   G(\vec{r},\vec{r}^{\,\prime})& = 0, \quad   \vec{r} \in \mathcal{E} \quad \text{or} \quad |\vec{r}|\to\infty. \nonumber
\end{alignat}
Then the formal solution $\Phi(\vec{r})$ of the Poisson equation can be established. According to the boundary conditions, the Green's function vanishes at the electrodes and infinity. Then the contribution of the electrodes is determined, where $\mathcal{E}$ represents all electrodes. Considering the interaction of the particles and the influence of $\sigma_{\mathcal{S}}$, the potential is
\begin{align}\label{eqn:potential}
	  \Phi(\vec{r},t) =   \frac{1}{\varepsilon_0} \sum_{i=1}^N q_i\, G(\vec{r},\vec{r}_i(t))+ \frac{1}{\varepsilon_0}\int_{\mathcal{S}} \sigma_{\mathcal{S}} \, G(\vec{r},\vec{r}^{\,\prime})\, {\rm{d}}f^{\,\prime} +  \int_{\mathcal{E}} \Phi(\vec{r}^{\,\prime}) \varepsilon_{\rm r}(\vec{r}^{\,\prime})  \nabla^\prime G(\vec{r},\vec{r}^{\,\prime})  \cdot\! {\rm{d}}\vec{f}^{\,\prime}. 
\end{align}

Containing the information about the geometry, the latter formula is defined as the \linebreak characteristic function $\Psi_{n}(\vec{r}\,)$, which is independent of the plasma
\begin{align} 
	  \Psi_{n}(\vec{r}\,) = \int_{\mathcal{E}}  \varepsilon_{\rm r}(\vec{r}^{\,\prime})  \nabla^\prime G(\vec{r},\vec{r}^{\,\prime})  \cdot\! {\rm{d}}\vec{f}^{\,\prime}. 
\end{align}

Then the electric potential $\Phi(\vec{r},t)$ inside $\mathcal{V}$ can be written as the interaction between a pair of particles in addition to the reaction of a particle to the applied voltages and surface charge~$\Phi_{\mathcal{S}}$, 
\begin{align} \label{eqn:solver}
	  \Phi(\vec{r},t) =  \frac{1}{\varepsilon_0}\sum_{i=1}^N q_i\, G(\vec{r},\vec{r}_i(t))+\Phi_{\mathcal{S}}(\vec{r}\,) + \sum_{n=1}^{2} u_n(t)  \Psi_{n}(\vec{r}).
\end{align}

The dynamics of plasma particles can be described in Hamiltonian, where the kinetic energy is obtained from the conjugate momentum $\vec{p}_k$ of free point charges with mass $m_k$ to the position $\vec{r}_k$
\begin{align}
\!\!\! H\!\left(\vec{r}_1,\ldots,\vec{r}_N,\vec{p}_1,\ldots,\vec{p}_N\right) = \sum_{k=1}^N \cfrac{\vec{p}_k^2}{2m_k} + V\left(\vec{r}_1,\ldots,\vec{r}_N\right).
\end{align}
The potential energy is determined from the electric potential in Eqn.~\eqref{eqn:solver}. Considering that each set of pair only interacts once in two-body interaction, it can be written as
\begin{align}
 V\left(\vec{r}_1,\ldots,\vec{r}_N\right)=\frac{1}{2\varepsilon_0}\sum_{k=1}^N \sum_{\substack{i=1 \\ i\neq k}}^N q_k q_i G(\vec{r_k},\vec{r_i})+\sum_{k=1}^N q_k\left(\Phi_{\mathcal{S}}(\vec{r}_k)+\sum_{n=1}^{2} u_n(t)  \Psi_{n}(\vec{r}_k)\!\right).
\end{align}
The proposed scheme consists of two modules, field solver and particle pusher. The electric field can be calculated as the field solver, and the canonical equations of motion play the role as particle pusher. The terms with the Green's function can be combined due to symmetry. Therefore, for a random particle $k$
\begin{align}
&\frac{{\rm{d}}\vec{r}_k}{{\rm{d}}t} = \frac{\partial H}{\partial \vec{p}_k} = \frac{\vec{p}_k}{m_k},	\\
&\frac{{\rm{d}}\vec{p}_k}{{\rm{d}}t} =-\frac{\partial H}{\partial \vec{r}_k} =-\frac{1}{\varepsilon_0}   \sum_{\substack{i=1 \\ i\neq k}}^N q_k q_i \nabla_k G(\vec{r_k},\vec{r_i}) -  q_k \nabla_k\Phi_{\mathcal{S}}(\vec{r}_k) - \sum_{n=1}^{2} u_n(t) \nabla_k \Psi_{n}(\vec{r}_k).
\nonumber
\end{align} 

In general, a Green's function can be expanded in a set of basis functions depending on the geometry of the model. For the IMRP, the expansion is an infinite series of spherical harmonics with indices $l\, (0\leq l < \infty)$ and $m\, (-l \leq m \leq l)$: 
\begin{align} 
G(r,\theta,\varphi,r^{\,\prime},\theta^{\,\prime},\varphi^{\,\prime})&= 
\sum_{l=0}^\infty\sum_{m=-l}^l R_l (r,r^{\,\prime})
Y^{\star}_{lm}(\theta^{\,\prime},\varphi^{\,\prime})Y_{lm}(\theta,\varphi).
\end{align}
To avoid the singularity when two particles are almost identical, which leads to an infinite amount of energy, some regularization to truncate the noise in this model is required. In fact, the field or particle exhibits many random interactions in the system. They tend to cancel each other out due to the spherical symmetry, which allows the simplification of the sum of Green's functions. The truncation is realized by a projection operator $\hat{\mathrm{P}}$ on subspaces
\begin{align}
&\frac{{\rm{d}}\vec{p}_k}{{\rm{d}}t} =-\frac{\partial H}{\partial \vec{r}_k} =-\hat{\mathrm{P}}\left( \frac{1}{\varepsilon_0} \sum_{\substack{i=1 \\ i\neq k}}^N q_k q_i \nabla_k G(\vec{r_k},\vec{r_i})  +  q_k \nabla_k\Phi_{\mathcal{S}}(\vec{r}_k) + \sum_{n=1}^{2} u_n(t) \nabla_k \Psi_{n}(\vec{r}_k)\right),
\end{align} 
Due to the linearity, it is mathematically correct to shift the projection operator. Hence, the truncated Green's function can be found to effectively simplified the calculations, which is presented in Chapter 3,
\begin{align}
&\frac{{\rm{d}}\vec{p}_k}{{\rm{d}}t} =-\frac{\partial H}{\partial \vec{r}_k} =- \frac{1}{\varepsilon_0}\sum_{\substack{i=1 \\ i\neq k}}^N q_k q_i \nabla_k \hat{\mathrm{P}}  G(\vec{r_k},\vec{r_i})  -  q_k \nabla_k\hat{\mathrm{P}}\Phi_{\mathcal{S}}(\vec{r}_k) - \sum_{n=1}^{2} u_n(t) \nabla_k \hat{\mathrm{P}}\Psi_{n}(\vec{r}_k).
\end{align}

In the kinetic scheme, the applied voltages at the electrodes are provided as the input of the simulation. As the response, the charges on the electrodes $Q_n$ can be determined according to Gauss's law
\begin{align}
Q_n &=-\int_{\mathcal{E}_n} \varepsilon_0 \varepsilon_{\rm r}(\vec{r} )   \nabla\Phi(\vec{r})\cdot\! {\rm{d}}\vec{f}.
\end{align}

\pagebreak

\section{Spectral kinetic scheme of ideal MRP}
The Green's function of the IMRP-plasma system ($R_\mathrm{E} \le |\vec{r}| < \infty$) can be obtained explicitly. The basic idea is to truncate the interactions with a short distance $|\vec{r}-\vec{r}^{\,\prime}|$ so that the coefficients in spherical harmonics can be determined. The azimuthal symmetry in the potential indicates that the coefficient $m=0$. Besides, the antisymmetry with respect to~$\theta$~$\to$~$\pi-\theta$ leads to the vanishing of the expansion with the even coefficient $l$. Therefore, depending on the required mode number, $l$ can be assigned, where the projection operator~$\hat{\mathrm{P}}$ is determined as the truncation of the infinite series.

The radial functions $R_l(r,r^{\,\prime})$ is given as
\begin{align}
 R_l(r, r^{\,\prime})=
    \begin{cases}
      R^{(d,p)}_{l}(r, r^{\,\prime}), &  r \le R<r^{\,\prime}\quad \text{or} \quad r^{\,\prime} \le R<r \,\,\, \\
       \\
      R^{(p,p)}_{l}(r, r^{\,\prime}), &  r, r^{\,\prime}>R 
    \end{cases}.
\end{align}
The configuration-dependent coefficients $g_l$ and $f_l$ are defined for a compact notation, and~$r_>$~($r_<$) is the greater (smaller) of $r$ (see Appendix~\ref{App.Green}) 
\begin{align}
R^{(d,p)}_{l}(r, r^{\,\prime})&=g_l\left(\frac{r_<^l}{r_>^{l+1}}-\frac{R_{\rm E}^{2l+1}}{r^{l+1} r^{\prime  l+1}} \right) ,\\
R^{(p,p)}_{l}(r, r^{\,\prime})&=\frac{1}{2l+1}\left(\frac{r_<^l}{r_>^{l+1}}-f_l\frac{R^{2l+1}}{r^{l+1} r^{\prime  l+1}}  \right) . \nonumber
\end{align}

Considering the symmetry of two electrodes in IMRP, the symmetric $\Psi_{\rm{sym}}(r)$ and antisymmetric $\Psi_{\rm{anti}}(\vec{r})$ parts of the characteristic function are defined. Similarly, the applied voltages at the electrodes are separated into a symmetric part $u_{\rm{sym}}$ and an asymmetric part~$u_{\rm{asy}}$, which writes
\begin{align}
         \Psi_{\rm{sym}}(r)   &  =\Psi_1(\vec{r}) +\Psi_2(\vec{r}),& \Psi_{\rm{anti}}(\vec{r})  &  =\frac{1}{2}\big(\Psi_1(\vec{r})-\Psi_2(\vec{r})\big) . \\  
          u_{\rm{sym}}(t)  &  =\frac{1}{2}\left(u_1(t)+u_2(t)\right),& u_{\rm{asy}}(t) & = u_1(t)-u_2(t). 
\end{align} 
The potential of surface charge $\Phi_{\mathcal{S}}(r)$ can be established in the form of Green's function, which is outlined in appendix~\ref{App.S}. For the Green's function in the static condition ($l=0$ and~$m=0$),~$\hat{\mathrm{P}}$ can be eliminated. Then the potential can be simplified as separated parts,
\begin{align} \label{eqn:modified_phi}
\Phi(\vec{r},t) =  \frac{1}{\varepsilon_0}\sum_{i=1}^N q_i\, \hat{\mathrm{P}} G(\vec{r},\vec{r}_i(t))+   \Phi_{\mathcal{S}}(r) + u_{\rm{sym}}(t)\Psi_{\rm{sym}}(r)+ u_{\rm{asy}}(t)\hat{\mathrm{P}}\Psi_{\rm{anti}}(\vec{r}).
\end{align}
The explicit expression of $\Psi_{\rm{sym}}(r)$, $\hat{\mathrm{P}}\Psi_{\rm{anti}}(\vec{r})  $, $u_{\rm{sym}}(t)$, and $ u_{\rm{asy}}(t)$ are required to determine a \linebreak mathematically compact equation for the potential. As is defined, characteristic functions~$\Psi(\vec{r})$ obey the Laplace equation in domain $\mathcal{V}$, which can be expanded in spherical harmonics. The general solutions in $r$-direction are
\begin{align}
 \psi_l(r) =
\begin{cases}
      a^{\mathcal{D}}_l \dfrac{r^l}{R^l}+
        b^{\mathcal{D}}_l \dfrac{R^{l+1}}{r^{l+1}}, &  R_{\rm E} \le r<R  \\
     \\
      b^{\mathcal{P}}_l \dfrac{R^{l+1}}{r^{l+1}}, &  R \le r < +\infty 
    \end{cases}.
\end{align}
According to the continuity of the vacuum potential, the electric flux density at the surface of the dielectric, and the boundary condition at the electrodes, the coefficients~$a^{\mathcal{D}}_l$,~$b^{\mathcal{D}}_l$,~and~$b^{\mathcal{P}}_l$ can be evaluated
\begin{align}
a^{\mathcal{D}}_l=\frac{(l+1)(\varepsilon_{\rm_r}-1)(1-d)^{(l+1)}}{(l+1)(\varepsilon_{\rm_r}-1)(1-d)^{(2l+1)}+l\varepsilon_{\rm_r}+l+1}, \nonumber  \\
b^{\mathcal{D}}_l=\frac{(1-d)^{(l+1)}(l\varepsilon_{\rm_r}+l+1)}{(l+1)(\varepsilon_{\rm_r}-1)(1-d)^{(2l+1)}+l\varepsilon_{\rm_r}+l+1},\\
b^{\mathcal{P}}_l=\frac{(2l+1)\varepsilon_{\rm_r}(1-d)^{(l+1)}}{(l+1)(\varepsilon_{\rm_r}-1)(1-d)^{(2l+1)}+l\varepsilon_{\rm_r}+l+1}. \nonumber
\end{align}
After $\Psi(\vec{r})$ is determined, it is possible to obtain $\Psi_{\rm{sym}}(r)$ and $\Psi_{\rm{anti}}(\vec{r})$ according to its definition. ${\Psi}_{\rm{sym}}(r)$ represents the static situation, the explicit form is
\begin{align}\label{eqn:symmetric influence function}
 \Psi_{\rm{sym}}(r)=
\begin{cases}
     \dfrac{1-d }{d+\varepsilon_{\rm_r}-d\varepsilon_{\rm_r}}\dfrac{R}{r}+ \dfrac{(1-d)(\varepsilon_{\rm_r}-1) }{d+\varepsilon_{\rm_r}-d\varepsilon_{\rm_r}}, &  R_\mathrm{E} \le |\vec{r}| < R   \\
       \\
      \dfrac{(1-d)\varepsilon_{\rm_r}}{d+\varepsilon_{\rm_r}-d\varepsilon_{\rm_r}}\dfrac{R}{r},  &  R \le |\vec{r}| < +\infty
    \end{cases}.
\end{align}

The antisymmetric function $\Psi_{\rm{anti}}(\vec{r})$ is expressed in the form of the Legendre series. The coefficient $c_l$ is determined by means of Rodrigues' formula \cite{Jackson2006}, which contains the \linebreak information about the electrode configuration within the probe tip. Due to the \linebreak antisymmetric excitation at the electrodes, all the terms with even $l$ are canceled out, and only the odd terms are different from zero. It is convenient to define ${l=2l^{\prime}-1 \, \forall \, l^{\prime}\in \mathbb{N}}$, then we obtain 
\begin{align}
\Psi_{\rm{anti}}(\vec{r})=\sum_{l^{\prime}=1}^{\infty}c_{l^{\prime}} \psi _{l^{\prime}}(r)P_{l^{\prime}}(\cos\theta) 
\quad\mbox{      with      }\quad c_{l^{\prime}}=\big(-\frac{1}{2}\big)^{{l^{\prime}}+1}\frac{(4{l^{\prime}}-1)(2{l^{\prime}}-3)!!}{{l^{\prime}}!} .
\end{align} 
The signal provided by the probe is related to $u_{\rm{sym}}(t)$ and $u_{\rm{asy}}(t)$. $u_{\rm{sym}}(t)$ is the floating potential, and $u_{\rm{asy}}(t)$ plays the role as the input of the simulation. The charge on the electrodes $Q_n$ is defined as the response of the system, where $Q_{\rm{tot}}(t)$ represents the total charge on the electrodes, and $Q_{\rm{diff}}(t)$ is the charge difference. 
When the probe is in the plasma without excitation, the floating potential $u_{\rm{sym}}(t)$ can be determined from $Q_{\rm{tot}}(t)$ in the static condition. After the probe is switched on, the perturbation occurs, which is captured in $Q_{\rm{diff}}(t)$.
\begin{align}
Q_{\rm{tot}}(t)   &  =Q_1(t)  + Q_2(t) ,&  Q_{\rm{diff}}(t) & =  \frac{1}{2}\big(Q_1(t) -Q_2(t) \big) . 
\end{align} 
According to Gauss's law, the charge on the electrode $Q_n(t)$ can be calculated, $C_{nn^{\prime}}
$ are defined as the capacitance coefficients, (The detailed calculation is shown in Appendix~\ref{App.ChargeQ}.) 
\begin{align} \label{eqn:Q_define}
Q_n(t) &=-\int_{\mathcal{E}_n} \varepsilon_0 \varepsilon_{\rm r}  \nabla\Phi(\vec{r})\cdot\! {\rm{d}}\vec{f} \nonumber \\
&= -\sum_{i=1}^N q_i  \Psi(\vec{r}_i(t)) -\int_{\mathcal{S} } \sigma_{\mathcal{S} }   \Psi_{n}(\vec{r}^{\,\prime})  {\rm{d}}f^{\,\prime} + \sum_{n^{\prime}=1}^{2} C_{nn^{\prime}} u_{n^{\prime}}(t).
\end{align}
To determine the floating potential, the case $r>R$ is solved. Then the total charge is obtained from the characteristic functions,
\begin{align}
 Q_{\rm{tot}}=-\frac{(1-d)R\varepsilon_{\rm r}}{d+\varepsilon_{\rm r}-d\varepsilon_{\rm r}}\sum^{N}_{i=1} \frac{q_i}{r_i}-Q_{\mathcal{S}}\dfrac{(1-d)\varepsilon_{\rm r}}{d+\varepsilon_{\rm r}-d\varepsilon_{\rm r}}+\frac{4\pi\varepsilon_{\rm r}\varepsilon_0(1-d)R}{d+\varepsilon_{\rm r}-d\varepsilon_{\rm r}}u_{\rm{sym}}.
\end{align}
The floating potential is calculated from $Q_{\rm{tot}}=0$,
\begin{align}\label{eqn:floating_u}
u_{\rm{sym}}=\frac{1}{4\pi\varepsilon_0}\left( \frac{Q_{\mathcal{S}}}{R}+ \sum_{i=1}^{N} \frac{q_i}{r_i} \right).
\end{align}
Inserting the expression of $u_{\rm{sym}}$, the potential in the domain $\mathcal{V}$ is
\begin{align} 
	  \Phi(\vec{r},t) =  \frac{1}{\varepsilon_0}\sum_{i=1}^N q_i\,\hat{\mathrm{P}} G(\vec{r},\vec{r}_i(t))+ \Phi_\mathcal{S}(r)+\frac{1}{4\pi\varepsilon_0}\left(\frac{Q_\mathcal{S}}{R}+ \sum_{i=1}^{N} \frac{q_i}{r_i} \right)\Psi_{\rm{sym}}(r)+ u_{\rm{asy}}(t)\hat{\mathrm{P}}\Psi_{\rm{anti}}(\vec{r}).
\end{align}
Then the modified field solver is completed with the influence of the surface charge $\Phi_{\mathcal{S}}(r)$ and the floating potential $u_{\rm{sym}}$. Focusing on the interaction within the plasma, the case~$r>R$ is considered. Therefore, we obtain
 \begin{align} 
\Phi_{\mathcal{S}}(r)+\frac{1}{4\pi\varepsilon_0}\frac{Q_{\mathcal{S}}}{R}\Psi_{\rm{sym}}(r)=\frac{Q_{\mathcal{S}}}{4\pi\varepsilon_0 r},
\end{align}
and a new modified Green's function can be defined as
\begin{align} 
\tilde{G}(\vec{r},\vec{r}^{\,\prime})=\hat{\mathrm{P}}  G(\vec{r},\vec{r}^{\,\prime})+\frac{1}{4\pi r'}\Psi_{\rm{sym}}(r).
\end{align}
Summing up all the related terms, the potential in the domain $\mathcal{V}$ is
 \begin{align} 
\Phi(\vec{r},t) =  \frac{1}{\varepsilon_0}\sum_{i=1}^N q_i\, \tilde{G}(\vec{r},\vec{r}_i(t)) + u_{\rm{asy}}(t) \hat{\mathrm{P}}\Psi_{\rm{anti}}(\vec{r}) +\frac{Q_{\mathcal{S}}}{4\pi\varepsilon_0 r}.
\end{align}

According to the results in \cite{LapkeMussenbrockBrinkmann2008,JensBrinkmann_Eigen2014}, the prominent feature of the resonance spectrum of the IMRP is the absorption peaks. The first absorption peak, the so-called dipole mode, is the dominant one. The resonances of higher modes are barely visible. Therefore, in addition to the static situation ($l=0$), considering only the dipole mode ($l=1$) in the simulation can be seen as an applicable approximation. Thus, the projection operator $\hat{\mathrm{P}}$ is determined, which leads to a simplified model with the truncated Green's function
 \begin{align} 
\tilde{G}(\vec{r},\vec{r}^{\,\prime})=\frac{1}{4\pi}\left( \frac{1}{r_>} +\frac{1}{4\pi}\left( \frac{r_<}{r^2_>}-f_1\frac{R^3}{r^2{r^{\,\prime}}^2}\right)\cos \theta\cos\theta' \right). 
\end{align}
Finally, the spectral kinetic scheme is defined where the signal provided by IMRP is $u_{\rm{asy}}$, and the response of the system is the charge difference on the electrodes $Q_{\rm{diff}}$ in the time domain. The spectral response of the IMRP-plasma system can be expressed by the real part of the admittance, which is determined from the Fourier transformation of $Q_{\rm{diff}}$.

\pagebreak

\section{Implementation and results}
\begin{figure}[!htb]
\centering
\includegraphics[width=0.9 \textwidth]{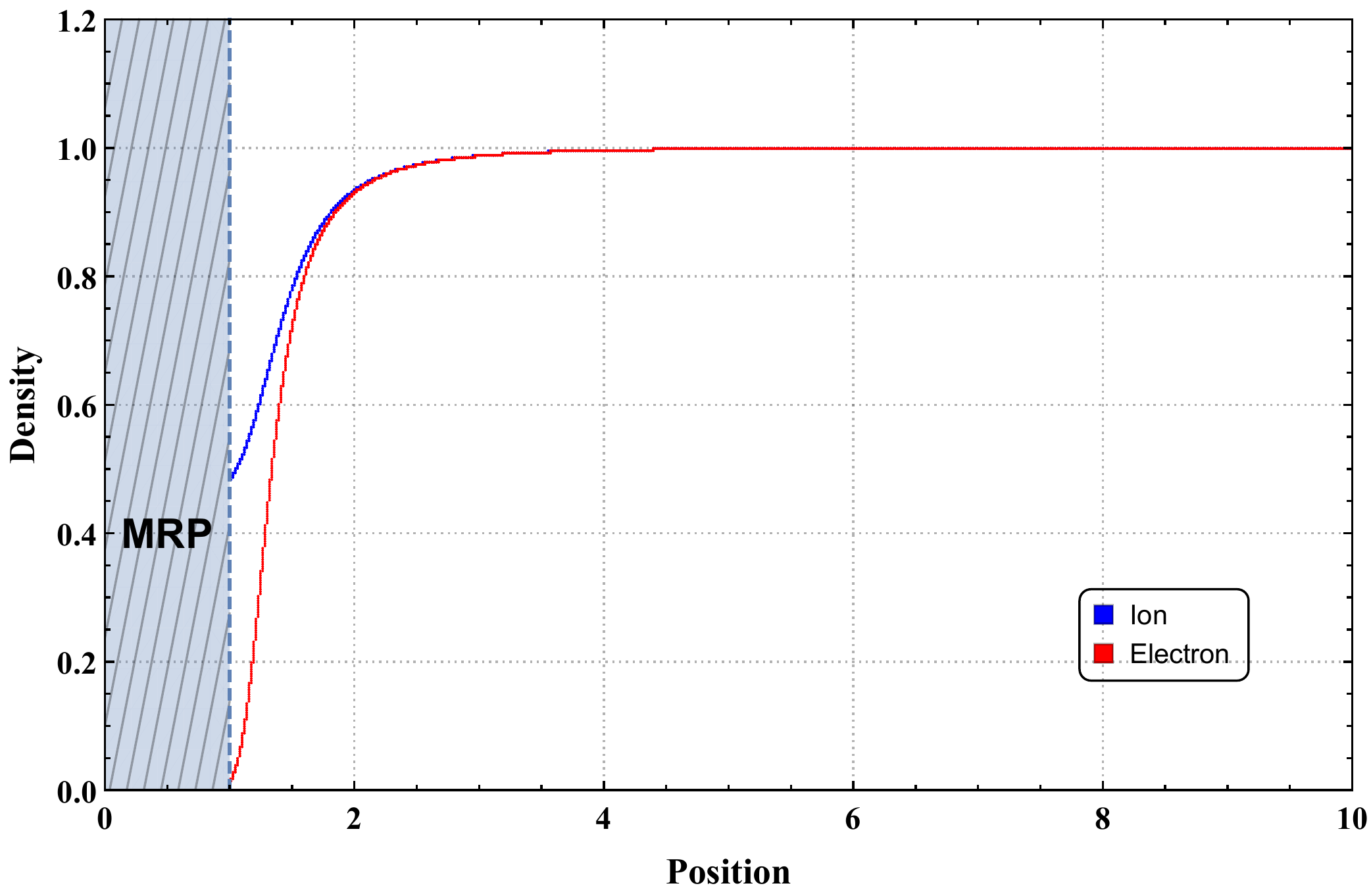} 
\caption{The density profile of the plasma-IMRP system in static condition}
\label{fig:Density}
\end{figure}

The radius of the IMRP is defined as $R=0.004$ $\rm{m}$, and the thickness of the dielectric is given by $Rd=0.001$ $\rm{m}$. It is useful to introduce dimensionless notation: $r\to Rr$, $t\to {\omega_{\rm{pe}}^{-1}}t$, $q\to eq$, $m\to m_{\rm{e}} m$, and $n\to n_{\infty} n$. The term $n_{\infty}$ is the electron density at infinity in the Poisson-Boltzmann equation, where the stationary state can be calculated as the initial condition of the simulation (see Appendix~\ref{App.Initial}). The investigated plasma is assumed with~$n_{\infty}=1~\times 10^{15}$ $\rm{m}^{-3}$~and~$T_{\rm{e}}=3$~$\rm{eV}$ in the simulation domain $r\in(1,10)$, and the system fulfills the charge equilibrium.  In Fig.~\ref{fig:Density}, the static density profiles for ions and electrons are presented, and the static negative surface charge can be determined. Then the particles are distributed according to these calculations.

Depending on the number of super-particles, a parameter $\tilde{n}$ is defined as the expected number of super-particles in a unit cube. Therefore, the equations of motion is obtaied, which are invariant against rescaling the number of particles to the super-particles 
\begin{align}
  \frac{{\rm{d}}\vec{r}_k}{\rm{d}t} & =\vec{v}_k,	\\
  m_k\frac{\rm{d}\vec{v}_k}{\rm{d}t} &= 
      -\frac{1}{\hat{n}} \sum_{\substack{i=1 \\ i\neq k}}^N q_k q_i\, \nabla_k \tilde{G}(\vec{r}_k,\vec{r}_i) 
- 	 q_k \frac{1}{\tilde{n}}  \nabla_k \frac{Q_S}{4\pi r_k}	 
	  - q_k u_{\rm{asy}}\nabla_k\Psi_{\rm{anti}}^{l=1}(\vec{r}_k).
\end{align}

Finally, a numerical algorithm is defined to describe the interaction of plasma around the IMRP. The output of the system linearly depends on the present and past value of the input~($t \leq t_0$) in such a causal system. $u_{\rm{asy}}=1$ is defined as the signal of the ideal MRP in the simulation, which will be only applied at the first step $t=t_0$. Then the signal from the probe disappears from $t>t_0$. It demonstrates a behavior similar to an impulse to all the particles. The response of the system is the charge difference on the electrodes,
\begin{align}
	  Q_{\rm{diff}}=  -\frac{1}{\tilde{n}}\sum_{i=1}^{N}q_i \Psi^{l=1}_{\rm{anti}}(\vec{r}_i ).	  
\end{align} 
 
The boundary condition is a vital part of the simulation. The particles can travel to the surface of the IMRP or the outer boundary. Here, the diffuse reflection of the particles is considered: The particle is reflected back to the simulation domain with the same speed but random direction once it reaches the boundary. Considering the ions are so much heavier, most of the particles that reach the boundary are the electrons. Moreover, the number of those electrons is limited due to the strong influence of the sheath. Therefore, this boundary condition can be seen as the ideal assumption, especially since the heating phenomena caused by the input signal from the probe are negligible.

In this example, the initial number of the super-particles is given as $1\times 10^6$, including ions and electrons. To reveal the pure kinetic effects, the collisionless dynamics of particles are computed, i.e., the collision frequency $\nu=0$. In reality, the energy distribution function can be complicated. To find the relation between the plasma parameters to the measured $\Delta\omega$, the simulation with all the generated particles follows the Maxwellian velocity distribution function as the general case. Then the signal from the probe causes the perturbation in the plasma. However, it is relatively small, and the simulation only describes an impulse response in a short time scale. Therefore, the electrons return to Maxwellian distribution at the end of the simulation. Eventually, the resonance response is recorded to evaluate the simulated $\Delta\omega$.

\begin{figure}[!htb]
\centering
\includegraphics[width=0.9\textwidth]{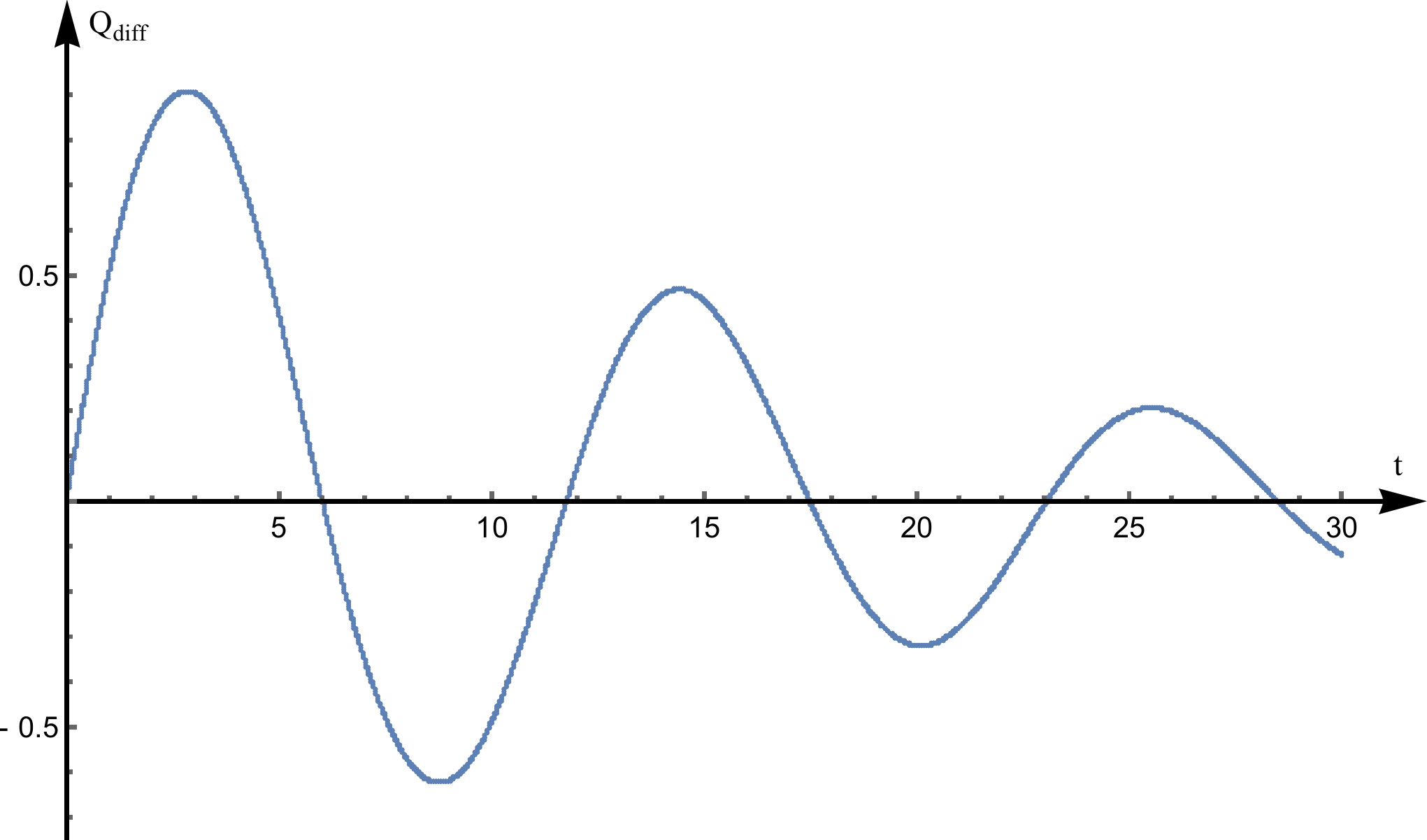}  
	\caption{Simulation result of the collisonless spectral simulation of ideal MRP: after the signal is applied via the probe, the charge difference $Q_{\rm{diff}}$ on the electrodes is recorded in the time domain, and the damped resonance is observed.}
	\label{fig:result1}
  \end{figure}

\begin{figure}[!htb]
\centering
 \includegraphics[width=0.9\textwidth]{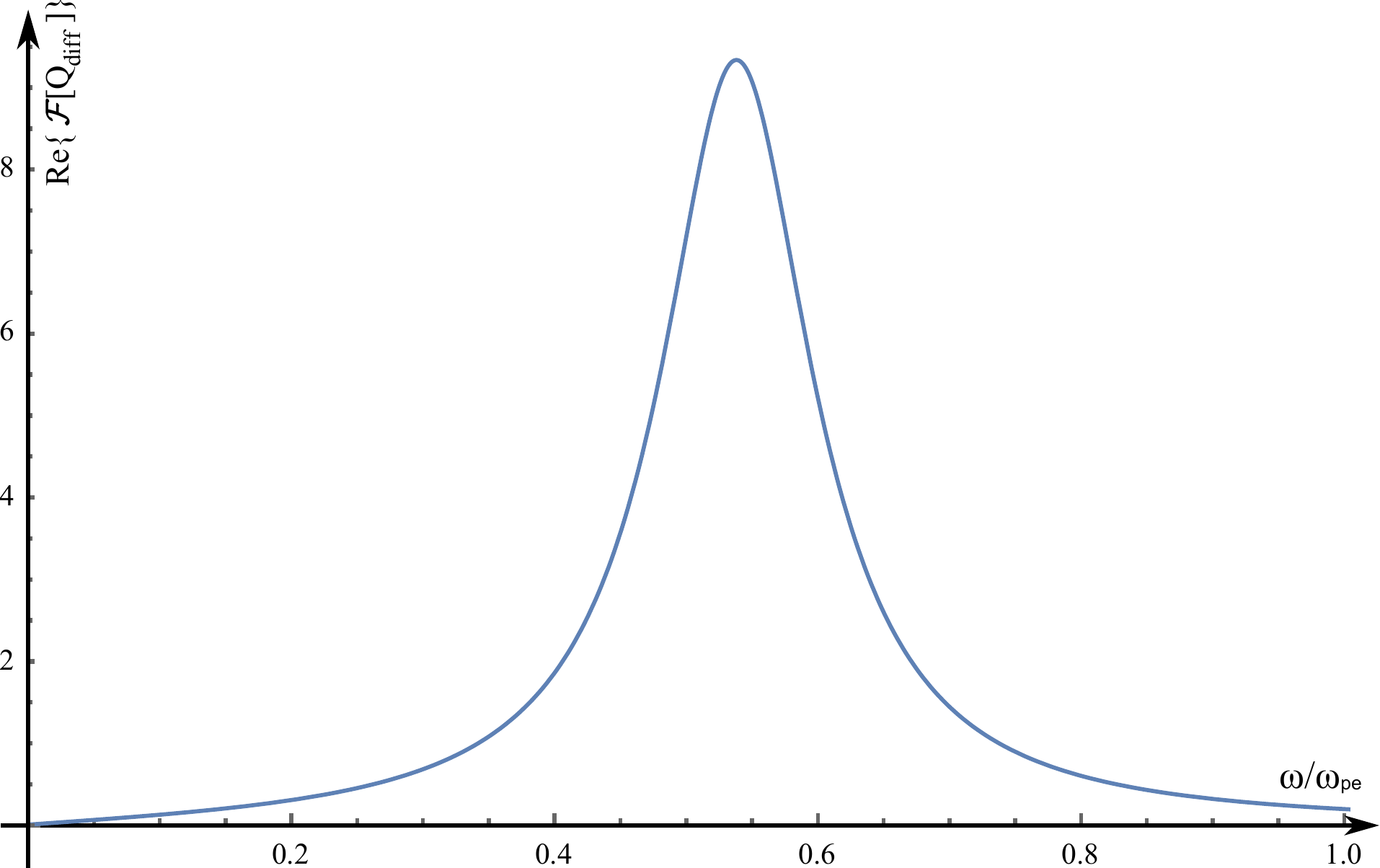}
	\caption{After the Fourier transformation of the charge difference $Q_{\rm{diff}}$, the resonance curve is presented in the frequency domain. The demonstrated resonance broadening is due to the kinetic effect.}
	\label{fig:result2}
  \end{figure}

Fig.~\ref{fig:result1} shows the charge difference on the electrodes at each time step ($\Delta t=0.01$) after the signal performs a ``kick'' to the plasma. The oscillation in the time domain is then observed. Moreover, the expected damping phenomenon is captured. It is convenient and insightful to analyze the continuous function. Therefore, the mathematical expression can be used to fit the numerical results  
\begin{align}
	  f(t)=\sum_{s=1}^{\infty} a_s e^{-b_s t} \sin(c_s t),
\end{align} 
where $a_s$ is the amplitude, $b_s$ is the damping factor and $c_s$ is the resonance frequency. In this example, the higher modes of the expansion are not necessary,  $s=1$ denotes an analytical function
\begin{align}
	  f(t)= 1.089 e^{-0.06317t} \sin(0.5352 t).
\end{align}
The Fourier transformation is derived from the fitting curve so that the result in the \linebreak frequency domain can be determined. Fig.~\ref{fig:result2} shows the corresponding spectrum of the IMRP: since the higher modes are ignored, the simulation of the dipole mode provides only the dominant resonance peak.

A similar resonance spectrum is also presented and analyzed in~\cite{LapkeMussenbrockBrinkmann2008,JensBrinkmann_Eigen2014}. In the Drude model, the MRP system is treated equivalently as an infinite number of series resonance circuits, each representing a resonance mode, parallel to a vacuum coupling $C_{\rm{vac}}$. The probe response is characterized by the complex admittance $Y$, where $C_l$ describes the capacitances of the resonance circuits and $\eta_l\omega_{\rm{pe}}$ indicates the resonance frequencies of the modes,
\begin{align}
	 Y(\omega)=i\omega C_{\rm{vac}}+\sum_{l=1}^\infty C_l \left( \frac{1}{i\omega}+\frac{i\omega+\nu}{\eta_l^2\omega_{\rm{pe}}^2} \right)  ^{-1}.
\end{align} 
The explicit values for the resonance in dipole mode ($l=1$) can be determined by assuming the sheath thickness and the collision frequency. It is noteworthy that the only energy loss in the Drude model within the plasma is due to the collisions. Therefore, it is necessary to set the value of the collision frequency in the Drude model.

\begin{figure}[!htb]
\centering
\includegraphics[width=0.9\textwidth]{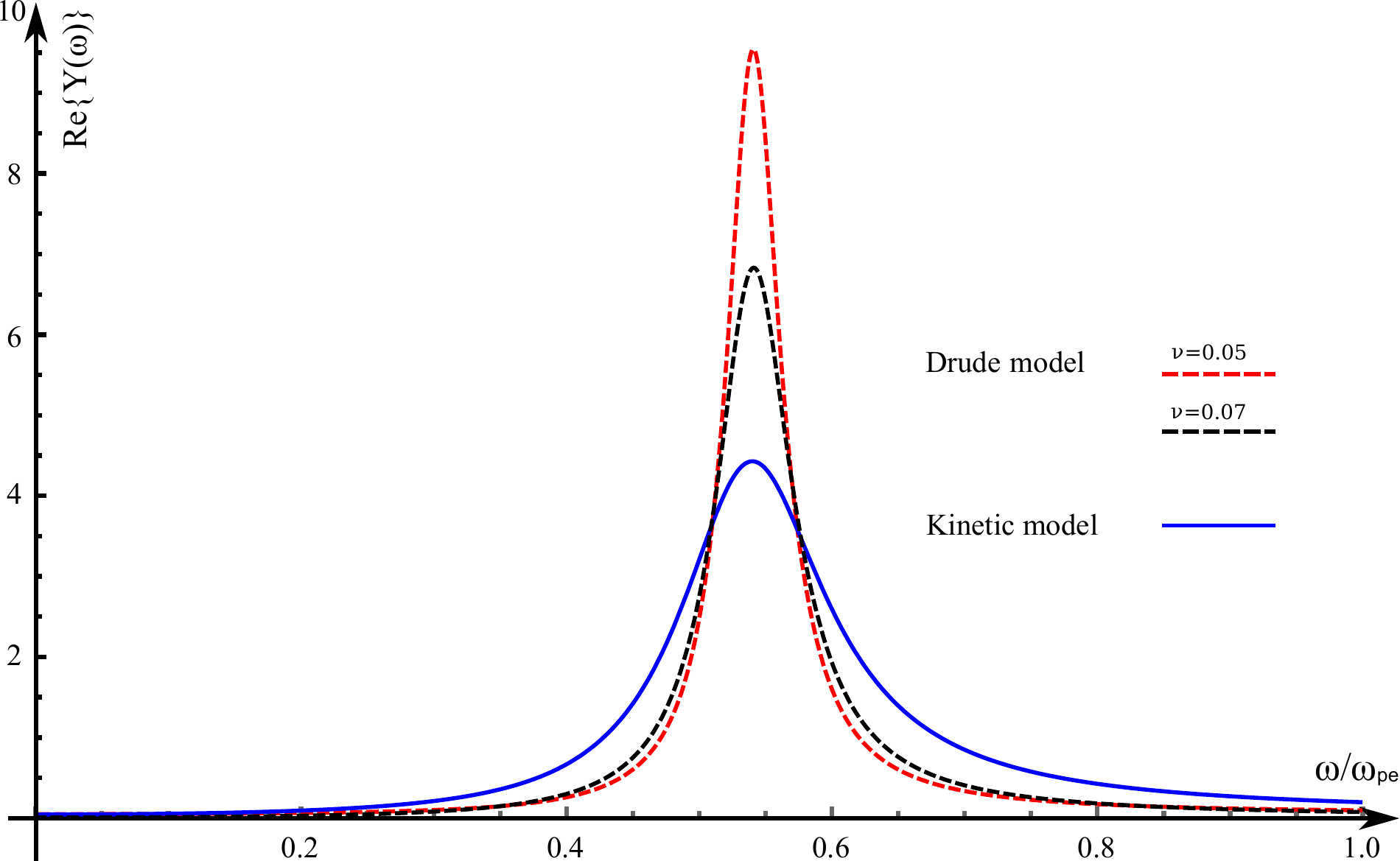}  
	\caption{Comparison of the resonance curves between the Drude model and the kinetic model: the admittance can be calculated directly in the Drude model with collision frequency at $0.05\omega_{\rm{pe}}$ (dashed red line) and $0.07\omega_{\rm{pe}}$ (dashed black line) whereas it is derived from the charge on the electrodes in the kinetic model (solid blue line).}
	\label{fig:admittance}
  \end{figure}

The comparison of the admittance between the kinetic model and the Drude model \linebreak is presented in Fig.~\ref{fig:admittance}. The admittance of the ideal MRP in the kinetic model can be derived from the simulated charge difference, and its spectrum is plotted. To have a clear view of the comparison, the sheath thickness is chosen accordingly in the Drude model to match the resonance frequency of the result in the kinetic simulation. In the comparison, an increasing collision frequency causes a broadening $\Delta \omega$ and a decreasing amplitude in the Drude model. However, even in the collisionless case, the resonance curve of the kinetic model is much broader, which indicates that the Drude model is with limited validity due to the absence of the kinetic effects. Contrarily, these effects are well demonstrated in the spectral kinetic simulation. The damped oscillation reflects the energy loss due to the escape of the free particles from the influenced domain.

\begin{figure}[!htb]
\centering
\includegraphics[width=0.9\textwidth]{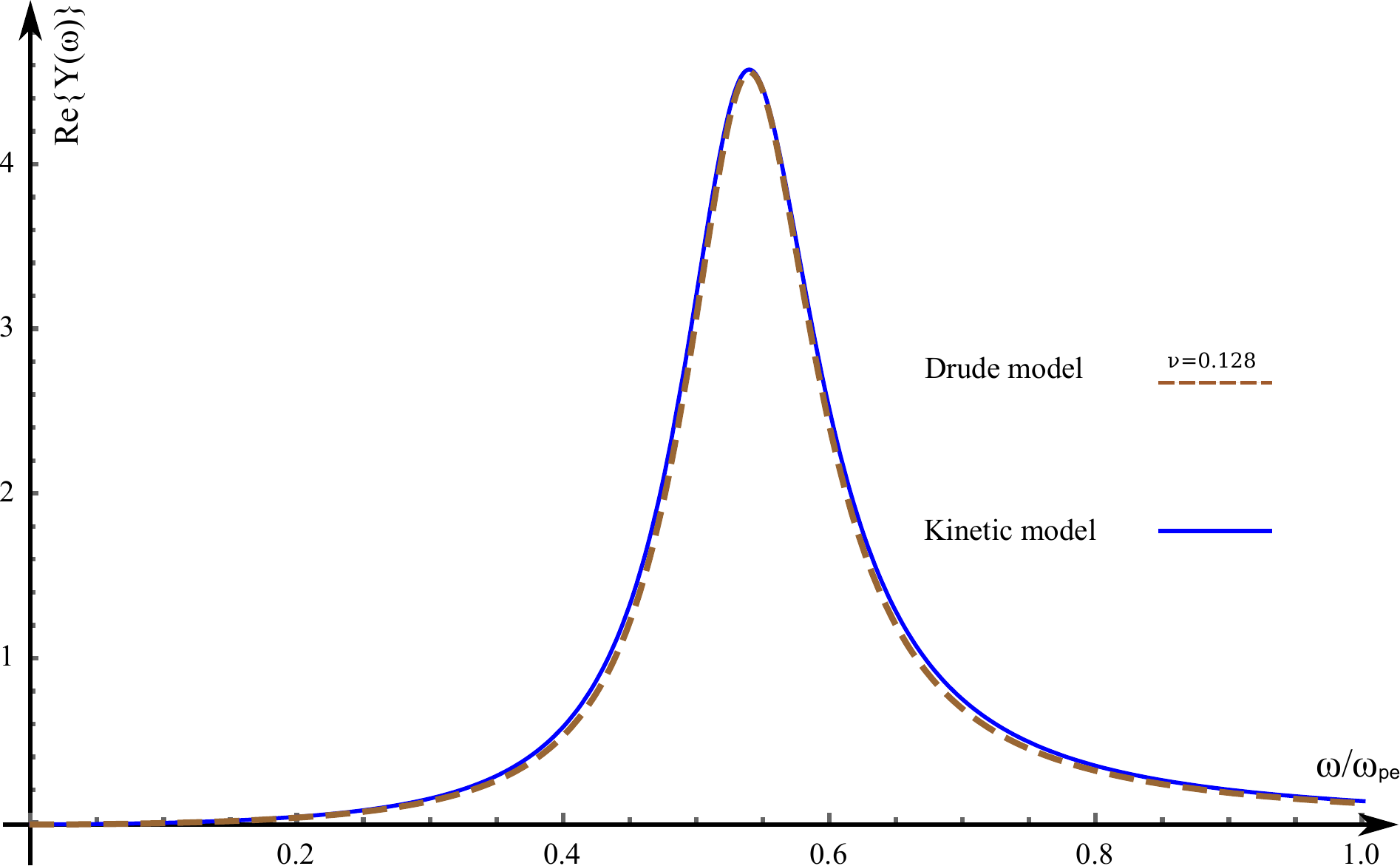}  
	\caption{Comparison of the resonance curves between the Drude model (dashed line) and the kinetic model (solid line): the effective collision rate $\nu_{\rm{eff}}$ can be determined by matching $\Delta \omega$ of the resonance peaks.}
	\label{fig:admittance_eff}
  \end{figure}

According to the analytical description of the MRP, $\Delta \omega$ of the mentioned resonance curve of the admittance is proportional to the effective collision rate $\nu_{\rm{eff}}$, which represents losses within the plasma. The electrons are deflected by the field of the MRP, which can be described as a collision rate~\cite{Lieberman2005,PopovGodyak1985}. It indicates that $\Delta \omega$ of the simulated resonance is of particular interest in order to evaluate $\nu_{\rm{eff}}$. With length scale $R$ and unknown coefficient $K$, assuming proportionality between the thermal velocity of the electrons $v_{\rm{th,e}}$ and the effective collision frequency $\nu_{\rm{eff}}$~\cite{SchulzStyrnollOberrath2013} allows such an expression
\begin{align}
	   \nu_{\rm{eff}}=K\cfrac{v_{\rm{th,e}}(T_{\rm{e}})}{R} = K\cfrac{\lambda_{\rm{D}}{\omega_{\rm{pe}}}}{R}.
\end{align} 
To eventually decide this $K$ to complete the proposed formula, $\nu_{\rm{eff}}$ is determined by \linebreak matching $\Delta \omega$ in the resonance peaks of the Drude model and the kinetic model. The admittance of the ideal MRP in the Drude model $Y_{\rm{Drude}}$ is introduced previously. It can be expressed as the function of the sheath thickness $\delta$, the collision rate $\nu$ and the frequency $\omega$ itself, whereas the admittance in the kinetic model $Y_{\rm{Kin}}$ is only the function of the frequency. The curve fitting can be implemented by the method of least squares, which writes
\begin{align}
 I=\int^{\omega_{\rm{pe}}}_{0} \left(Y_{\rm{Drude}}(\delta,\nu,\omega)-Y_{\rm{Kin}}(\omega)\right)^2 \rm{d}\omega.
\end{align} 
Therefore, finding the minimal value of $I$ provides us the $\nu_{\rm{eff}}$ from the evaluated $\nu$ in the Drude model. As shown in Fig.~\ref{fig:admittance_eff}, the curves match at $\nu=0.128\omega_{\rm{pe}}$. In this case, the coefficient $K=1.26$ is calculated. 

\begin{figure}[!htb]
\centering
\includegraphics[width=0.9\textwidth]{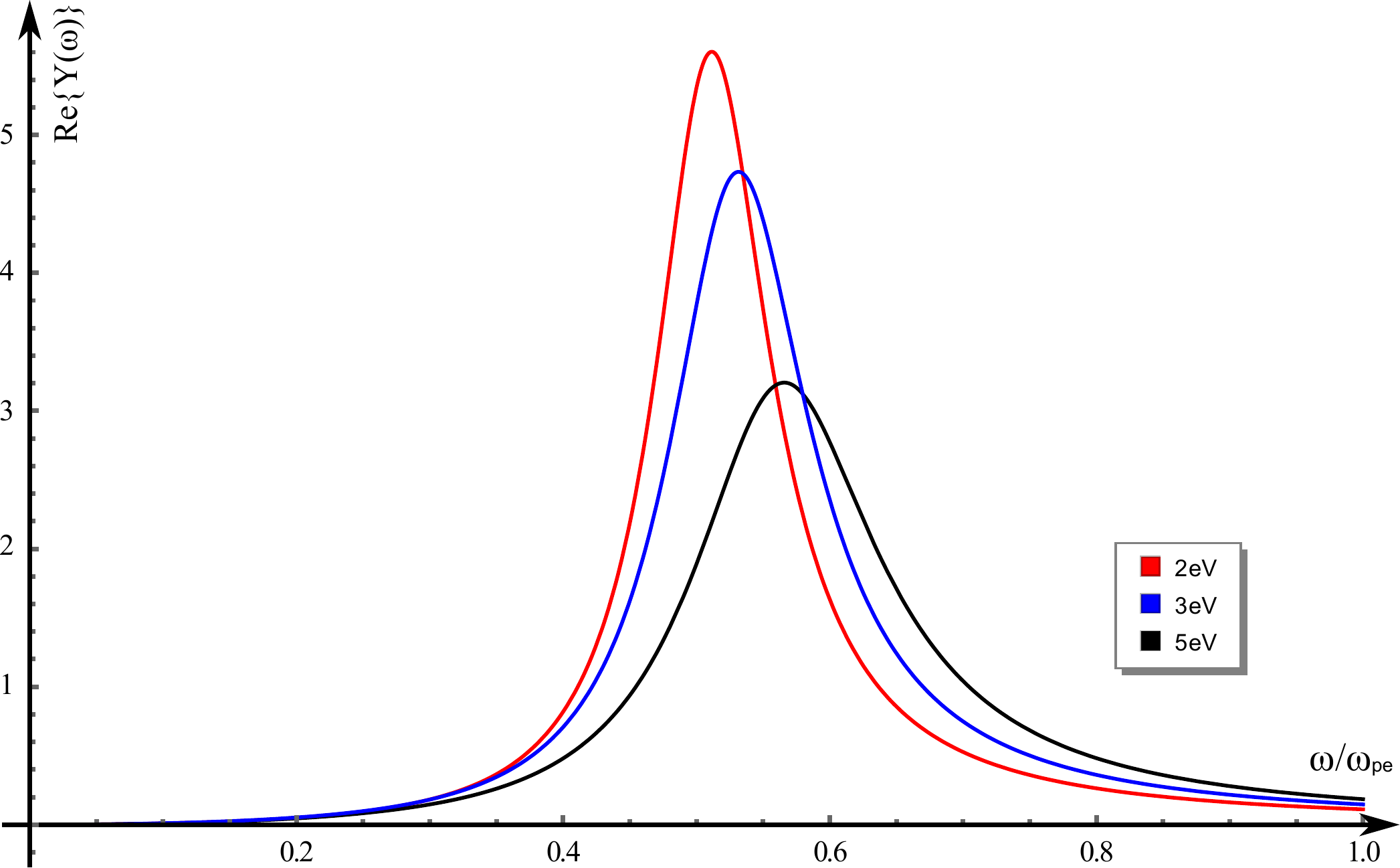}  
	\caption{Simulated resonance behavior of the IMRP with different electron temperature}
	\label{fig:diff_Te}
 \end{figure}

Similarly, the simulation for a variation of the electron temperature~$T_{\rm{e}}$~$\in$~\{2,~3,~5\}~$\rm{eV}$ is shown in Fig.~\ref{fig:diff_Te}. The effective collision rate is determined numerically, where $\nu_{\rm{eff}}$ $\in$ \{0.105, 0.128, 0.167\}$\omega_{\rm{pe}}$, the corresponding $K$ is then obtained for each electron temperature. The increase of the electron temperature results in the broadening of the resonance curves, which is related to the kinetic effects. Consequently, the coefficient $K=1.264$ is obtained in this specific example.

\pagebreak

\section{Summary and conclusion}
In this work, a kinetic scheme for a plasma-probe (MRP) system is derived. The subject of this investigation is the interaction of the probe with the plasma within its influence domain $\mathcal{V}$. The kinetic model of an idealized version of the MRP is presented, which gives physical insight into the damped resonance behavior. In the ideal case, the explicit form of the potential is defined and presented by an analytic expression. The Green's function is given by an infinite expansion and has to be truncated to determine a specific spectrum of ideal MRP. Therefore, only the dipole mode is taken into consideration. The higher modes are absent due to their minor influence on the results.

As an example, the motion of all the particles is simulated after applying a signal from the ideal MRP. The charge difference on the electrodes is recorded as the output of the simulation, and it is analyzed in the time and frequency domain. A comparison between the kinetic model and the Drude model is presented. Notably, the damping phenomenon and the broadened resonance curve in the kinetic model are obtained as expected. We define these as the kinetic effects which cover the energy loss due to the escape of the free particles from the influenced domain. Hence, the essence of this kinetic scheme is well demonstrated. 

Since the resonance frequency $\omega_{\rm{r}}$ is proportional to the plasma frequency $\omega_{\rm{pe}}$ and the relation of $\Delta \omega$ to the Debye length $\lambda_D$ can also be determined, the presented kinetic model of ideal MRP provides us the possibility to obtain the electron density and the electron temperature simultaneously from the simulated resonance curve. In the further study, the validation of the spectral kinetic scheme is to be resolved. Therefore, a parameter study of different $T_{\rm{e}}$ and $n_{\rm{e}}$ is necessary. Collisions between electrons and neutral atoms will be included in the kinetic model to compare the results with the measurements. Additionally, the assumption of different energy distribution functions will be studied. We are optimistic that these mentioned aspects will be implemented and discussed in detail in future work.

\newpage
\section{Citation}

\newpage

\begin{appendix}

\section{The Green's function for the ideal MRP}
\label{App.Green} 
In electrostatics, the Green's function $G(\vec{r},\vec{r}^\prime)$ is the 
solution of Poisson's equation with specified boundary conditions for a 
unit charge at the point $\vec{r}^\prime$. For the case of the ideal MRP,\linebreak the problem is to find the potential outside of a sphere of radius $R_\mathrm{E}$ that is covered with a dielectric of thickness $d$ and permittivity $\varepsilon_\mathrm{r}$ 
so that the total device radius is $R=R_\mathrm{E}+Rd$. \linebreak It is assumed that the sphere is grounded, and that the potential vanishes at infinity:
\begin{align}\label{eqn:Appendix Poisson}
    - &\nabla  \cdot\! \left(\varepsilon_{\rm r}(\vec{r} ) \nabla  G(\vec{r},\vec{r}^{\,\prime})\right)
	= \delta^{(3)}(\vec{r}\,-\vec{r}^{\,\prime}),\;\; \\ \nonumber
   &G(\vec{r},\vec{r}^{\,\prime})
   = 0 \;\; \text{for $|\vec{r}| = R_\mathrm{E}$
        or $|\vec{r}| \to \infty$}.
\end{align}
The function $\varepsilon_{\rm r}(\vec{r})$ describes the dielectric cover of the probe,
\begin{align} 
	    \varepsilon_{\rm r}(\vec{r})  = \begin{cases}
	         \varepsilon_{\rm{r}}, & R_\mathrm{E} \le |\vec{r}| < R,\\
	         1, & R \le |\vec{r}| < \infty, 
	    \end{cases}
\end{align}
It is, in fact, this cover that complicates the problem considerably.
If it were absent, $Rd = 0$, the solution could
be obtained by the mirror
principle~\cite{Jackson2006}:
\begin{align}
     G^{(Rd=0)}(\vec{r},\vec{r}^{\,\prime}) 
	= \cfrac{1}{ 4\pi \lvert \vec{r}\,-\vec{r}^{\,\prime} \rvert}
	-\cfrac{R_\mathrm{E}}{ 4\pi r^{\,\prime}  \lvert \vec{r}\,-({R_\mathrm{E}^2}/{{r^{\,\prime}}^2})\vec{r}^{\,\prime} \rvert}.
\end{align}
Here, however, a closed solution is not possible, and we aim instead for a series solution. \linebreak 
Obviously, because of symmetry, an expansion into spherical harmonics 
$Y_{lm}(\theta,\varphi)$ is possible, 
so we make the following ansatz, where the 
$R_l(r,r^{\,\prime})$ are yet unknown functions,
\begin{align}\label{eqn:Green Spherical}
G(r,\theta,\varphi,r^{\,\prime},\theta^{\,\prime},\varphi^{\,\prime})&= 
\sum_{l=0}^\infty\sum_{m=-l}^l R_l(r,r^{\,\prime})
Y^{\star}_{lm}(\theta^{\,\prime},\varphi^{\,\prime})Y_{lm}(\theta,\varphi).
\end{align}
The completeness relation is used to describe the delta function as
\begin{align}\label{eqn:completeness}
\delta^{(3)}(\vec{r}\,-\vec{r}^{\,\prime})= 
\frac{1}{r^2}\delta(r-r')\sum_{l=0}^\infty\sum_{m=-l}^l  
Y^{\star}_{lm}(\theta^{\,\prime},\varphi^{\,\prime})Y_{lm}(\theta,\varphi).
\end{align}

Inserting \eqref{eqn:completeness} and \eqref{eqn:Green Spherical} into \eqref{eqn:Appendix Poisson} leads to
\begin{align}\label{eqn:slope}
\frac{\partial }{\partial r }r^2 
\varepsilon_\mathrm{r}(r)\frac{\partial{ R_l(r, r^{\,\prime})}}{\partial r} -\varepsilon_\mathrm{r}(r)l(l+1)R_l(r, r^{\,\prime})=-\delta(r-r^{\,\prime}).
\end{align}
In intervals which contain neither $R$ nor $r^\prime$, 
the solutions are of the form
\begin{align}\label{eqn:general solution}
R_l(r, r^{\,\prime}) =    A(r^{\,\prime}) r^l +  B(r^{\,\prime}) \dfrac{1}{r^{l+1}}.
\end{align}

According to the geometry of the IMRP, the regional radial Green's function is defined in the range $R_\mathrm{E}\le|\vec{r}| < \infty$. $R^{(p,p)}_{l}$ describes the situation that $\vec{r}$ and $\vec{r}^{\,\prime}$ are both located in the plasma bulk whereas $R^{(d,p)}_{l}$ indicates that one of $\vec{r}$ or $\vec{r}^{\,\prime}$ is in the dielectric and the other is in the plasma bulk. Although charges only exist in the plasma bulk ($|\vec{r}^\prime| \ge R$), to solve the discontinuity at $\vec{r}=R$, the case for $R_\mathrm{E} \le|\vec{r}^\prime| < R$ in $R^{(d,p)}_{l}$ is required,
\begin{align} 
 R_l(r, r^{\,\prime})=
    \begin{cases}
      R^{(d,p)}_{l}(r, r^{\,\prime}), &  \text{$r \le R<r^{\,\prime}$ or $r^{\,\prime} \le R<r$}  \\
      R^{(p,p)}_{l}(r, r^{\,\prime}), &  r, r^{\,\prime} \in (R,+\infty)  
    \end{cases}. 
\end{align}

Firstly, $R^{(d,p)}_{l}(r, r^{\,\prime})$ is to be discussed. The coefficients $A(r^{\,\prime})$, $B(r^{\,\prime})$, $A'(r^{\,\prime})$ and $B'(r^{\,\prime})$ are to be determined according to the boundary conditions, which leads to the vanishing of~$R^{(d,p)}_{l}(r, r^{\,\prime})$ at infinity and $r=R_{\rm E}$,
\begin{align} 
 R^{(d,p)}_{l}(r, r^{\,\prime})=
    \begin{cases}
      A(r^{\,\prime})\left(r^l-\dfrac{R_{\rm E}^{ 2l+1 }}{r^{l+1}}\right), &  R_{\rm E}\le r<R\le r^{\,\prime}< +\infty  \\[2.0ex]
        B'(r^{\,\prime}) \dfrac{1}{r^{l+1}}, &  R_{\rm E}\le r^{\,\prime}<R\le r < \infty
    \end{cases}.
\end{align}
The symmetry of $R^{(d,p)}_{l}(r, r^{\,\prime})$ in $r$ and $r^{\,\prime}$ requires the coefficients $A(r^{\,\prime})$ and $B'(r^{\,\prime})$ be such that $R^{(d,p)}_{l}$ can be written  
\begin{align} 
R^{(d,p)}_{l}(r, r^{\,\prime})&=g_l\left(\frac{r_<^l}{r_>^{l+1}}-\frac{R_{\rm E}^{ 2l+1 }}{r^{l+1} r^{\prime  l+1}} \right)
\end{align}
where $r_<$ ($r_>$) represents the smaller (larger) of $r$ and $r^{\,\prime}$. The effect of the delta function is considered according to equation~\eqref{eqn:slope} to determine the constant $g_l$, where a discontinuity exists at $r=r^{\,\prime}=R$. It is integrated over the interval from $r=r'-\epsilon$ to $r=r'+\epsilon$, where $\epsilon$ is assumed to be a small number
\begin{align} \label{eqn:discontinuity}
\varepsilon_{\rm r}(\vec{r} ) \left.\frac{\rm{d}}{\rm{d}r}[ R^{(d,p)}_{l}(r,  r^{\,\prime})]\right\vert_{r'-\epsilon}^{r'+\epsilon}=-\frac{1}{ r^2 }.
\end{align}
Taking different permittivities into account
\begin{align}
g_l\left(-(l+1)\frac{r'^l}{ r^{ l+2 }}+(l+1)\frac{R_{\rm E}^{ 2l+1 }}{r^{ l+2 }r'^{l+1}}  \right) - g_l \epsilon_{\rm r}\left(l\frac{r^{l-1}}{r'^{l+1}}+ (l+1)\frac{R_{\rm E}^{2l+1}}{r^{ l+2 }r'^{l+1}}  \right)=-\frac{1}{r^2 }.
\end{align}
Therefore, we obtain  
\begin{align}
g_l = \frac{R ^{ 2l+1 }}{(l+1+l\varepsilon_{\rm r} )R ^{ 2l+1 }+(l+1)(\varepsilon_{\rm r}-1)R_{\rm E}^{ 2l+1 }}.
\end{align}
Then, for $r, r'>R$, the radial functions $R^{(p,p)}_{l}$ are discussed in the following. The boundary condition at infinity applies in the general form~\eqref{eqn:general solution}, besides, the jump conditions at $r=R$ between $R^{(d,p)}_{l}$ and $R^{(p,p)}_{l}$ can determine the coefficients of the radial functions $R^{(p,p)}_{l}$, which are written as
\begin{align}
 R^{(p,p)}_{l} \bigg\vert_{r\rightarrow \infty}  &  =0 ,& \varepsilon_{\rm r} \frac{\partial{ R^{(d,p)}_{l}}}{\partial{r}}  \bigg\vert_{r=R} & = \frac{\partial{ R^{(p,p)}_{l}}}{\partial{r}}  \bigg\vert_{r=R} . \nonumber
\end{align} 

Similarly, the discontinuity at $r=r'$ and the symmetry at $r$ and $r'$ lead to a simplified expression of $R^{(p,p)}_{l}$:  
\begin{align} 
R^{(p,p)}_{l}(r, r^{\,\prime})&=\frac{1}{2l+1}\left(\frac{r_<^l}{r_>^{l+1}}-f_l\frac{R ^{ 2l+1 }}{r^{l+1} r^{\prime  l+1}} \right), 
\end{align}
where
\begin{align} 
f_l=\frac{l(\varepsilon_{\rm r}-1)R ^{ 2l+1 }+(l+\varepsilon_{\rm r}+l\varepsilon_{\rm r})R_E ^{ 2l+1 }}{(l+1+l\varepsilon_{\rm r} )R ^{ 2l+1 }+(l+1)(\varepsilon_{\rm r}-1)R_{\rm E}^{ 2l+1 }}.
\end{align}
 
Consequently, the radial function $R_l(r, r^{\,\prime})$ of IMRP can be calculated explicitly, the configuration-dependent coefficients $f_l$ and $g_l$ in the radial functions are defined for a \linebreak compact notation.

\pagebreak
\section{Potential of the surface charge}\label{App.S}
The surface charge can be considered in a static situation, which leads to $l=0$ and $m=0$ in Green's function. The potential of the surface charges is written as:
\begin{align}
 \Phi_\mathcal{S}(r)=\frac{1}{\varepsilon_0}\int_{\mathcal{S}}\sigma_\mathcal{S} \, G(\vec{r},\vec{r}^{\,\prime})\, {\rm{d}}f^{ \,\prime}  = \frac{Q_\mathcal{S}}{4\pi \varepsilon_0 R^2}\int_{\mathcal{S}} G(\vec{r},\vec{r}^{\,\prime})\, {\rm{d}}f^{\, \prime}.
\end{align}
For the case $r<R$, according to the assumption, $r'\to R$ is applied, then $ R^{(d,p)}_{l}(r, r^{\,\prime})$ can be used to determine the solution:  
\begin{align}
 \left.\Phi_\mathcal{S}(r)\right\vert_{{r<R}} & = \frac{Q_\mathcal{S}}{4\pi \varepsilon_0 } R^{(d,p)}_{l=0}(r, r^{\,\prime})\nonumber \\
 &=  \frac{Q_\mathcal{S}}{4\pi \varepsilon_0 }\left(\frac{R}{R+(\varepsilon_{\rm r}-1)R_{\rm E} }\cdot\frac{1}{R} -\frac{R}{R+(\varepsilon_{\rm r}-1)R_{\rm E}}\cdot\frac{R_{\rm E}}{rR} \right) \\
 &= \frac{Q_\mathcal{S}}{4\pi \varepsilon_0 }\left(\frac{1}{R(d + \varepsilon_{\rm r}- d\varepsilon_{\rm r} )  }   -\frac{1-d}{r(d + \varepsilon_{\rm r}- d\varepsilon_{\rm r}  )} \right). \nonumber
\end{align}

Similarly, for the case $r>R$, the Greens function is in the form of $R^{(p,p)}_{l=0}(r, r^{\,\prime})$, we obtain 
\begin{align}
 \left.\Phi_\mathcal{S}(r)\right\vert_{{r>R}} & = \frac{Q_\mathcal{S}}{4\pi \varepsilon_0 } R^{(p,p)}_{l=0}(r, r^{\,\prime})\nonumber \\
 &=  \frac{Q_\mathcal{S}}{4\pi \varepsilon_0 }\left(\frac{1}{r}  -\frac{\varepsilon_{\rm r} R_{\rm E}}{R+(\varepsilon_{\rm r}-1)R_{\rm E}}\cdot\frac{R}{rR} \right) \\
 &= \frac{Q_\mathcal{S}}{4\pi \varepsilon_0 }\left( \frac{d}{r (d +R\varepsilon_{\rm r}- d\varepsilon_{\rm r} )}\right). \nonumber
\end{align}

\pagebreak
\section{Calculation of the charge on the electrodes}\label{App.ChargeQ}
The charge on the electrodes can be calculated according to Gauss's law:
\begin{align}\label{eqn:charge on electrodes}
Q_n=-\int_{\mathcal{E}_n} \varepsilon_0 \varepsilon_{\rm r}(\vec{r} )   \nabla\Phi(\vec{r})\cdot\! {\rm{d}}\vec{f}. 
\end{align}
Since the potential is derived, we can insert \eqref{eqn:potential} into \eqref{eqn:charge on electrodes}:
\begin{align} 
Q_n=-\int_{\mathcal{E}_n}   \varepsilon_{\rm r}(\vec{r} )   \nabla \left(
\sum_{i=1}^N q_i\, G(\vec{r},\vec{r}_i(t))
+ \int_{{\mathcal{S} }}\sigma_{\mathcal{S} } \, G(\vec{r},\vec{r}^{\,\prime})\, {\rm{d}}f^{\,\prime}\right. \nonumber \\
\left.+\int_{\mathcal{E}} \Phi(\vec{r}^{\,\prime})\varepsilon_0 \varepsilon_{\rm r}(\vec{r}^{\,\prime})  \nabla^\prime G(\vec{r},\vec{r}^{\,\prime})  \cdot\! {\rm{d}}\vec{f}^{\,\prime} 
\right)\cdot\! \rm{d}\vec{f}.
\end{align}
The influence of surface charge on the electrodes is defined as $Q_{{\mathcal{E}}_n}^{\mathcal{S}}$, and the equation can be simplified by replacing the characteristic function as
\begin{align}
	 Q_n(t) = -\sum_{i=1}^N q_i  \Psi(\vec{r}_i(t))+ Q_{{\mathcal{E}}_n}^{\mathcal{S}} + \sum_{n^{\prime}=1}^2 C_{nn^{\prime}} u_{n^{\prime}}(t),
\end{align}   
where capacity coefficient $C_{nn^{\prime}}$, which is proved symmetric, can be written in the following
\begin{align}
 C_{nn^{\prime}} &=  - 
		\varepsilon_0\int_{\mathcal{E}_n}\int_{\mathcal{E}_{n^{\prime}}} \varepsilon_{\rm r}(\vec{r}\,) \varepsilon_{\rm r}(\vec{r}^{\,\prime})\nabla \nabla^\prime G(\vec{r},\vec{r}^{\,\prime})\!\cdot\! \rm{d}\vec{f}^\prime\!\cdot\! {\rm{d}}\vec{f}  \nonumber \\
	&=	- 
		 \int_{\mathcal{E}_{n^{\prime}}} (\varepsilon_0 \varepsilon_{\rm r}(\vec{r}\,)  \nabla\Psi_n (\vec{r}\,))   \!\cdot\! {\rm{d}}\vec{f}^\prime\! \\
		 &=	- \int_{\mathcal{V}} (\varepsilon_0 \varepsilon_{\rm r}(\vec{r}\,)  \nabla\Psi_n (\vec{r}\,)   \!\cdot\! \nabla\Psi_{n^{\prime}} (\vec{r}\,))  {\rm{d}}^3r . \nonumber
\end{align} 
To be more specific for IMRP, the charge on each electrode is
\begin{align} 
Q_1 &= -\sum_{i=1}^N q_i  \Psi_1(\vec{r}_i(t))+\frac{1}{2}Q_{\mathcal{E}}^{\mathcal{S}}+ (C_{11}+C_{12})u_{\rm{sym}}(t)+(C_{11}-C_{12})\frac{ u_{\rm{asy}}(t)}{2},\\
Q_2 &= -\sum_{i=1}^N q_i  \Psi_2(\vec{r}_i(t))+\frac{1}{2}Q_{{\mathcal{E}}}^{\mathcal{S}} + (C_{21}+C_{22})u_{\rm{sym}}(t)+(C_{21}-C_{22})\frac{u_{\rm{asy}}(t)}{2},
\end{align}
where the influence of surface charge on both electrodes can be determined as
\begin{align} 
  Q_{\mathcal{E}}^{\mathcal{S}} &= -\int_{\mathcal{S} } \sigma_{\mathcal{S} }  (\Psi_{1}(\vec{r})+\Psi_{2}(\vec{r})) \rm{d}f \nonumber \\ &= -\int_{\mathcal{S} } \sigma_{\mathcal{S} }  \Psi_{\rm{sym}}(r) \rm{d}f \\
  &=-Q_{\mathcal{S}}\dfrac{(1-d)\varepsilon_{\rm r}}{d+\varepsilon_{\rm r}-d\varepsilon_{\rm r}}. \nonumber
\end{align}
Then the floating potential $u_{\rm{sym}}$ can be calculated by assuming the total charge on the electrodes in static situation as zero ($Q_{\rm{tot}}   =   Q_1+Q_2 = 0$),
\begin{align} 
u_{\rm{sym}}=\frac{1}{4\pi\varepsilon_0}\left( \frac{Q_{\mathcal{S}}}{R}+ \sum_{i=1}^{N} \frac{q_i}{r_i} \right).
\end{align}

\pagebreak

\section{Initial condition of the simulation}
\label{App.Initial}
The initial condition of the simulation, i.e., the state before the onset of the electric signal,\linebreak corresponds to a spherically symmetric probe-plasma equilibrium under  floating conditions. The
Poisson equation relates 
the potential $\Phi(r)$ to the ion and electron charge densities:
\begin{align}\label{eqn:app_Poisson}
      - \varepsilon_0\frac{1}{r^2}\frac{\partial}{\partial r} r^2  \frac{\partial\Phi}{\partial r} = e(n_{\rm{i}}(r)-n_{\rm{e}}(r)).
\end{align}
The electron density is described by the Boltzmann relation. Denoting the particle density far away from the probe (where the plasma is quasi-neutral) by $n_{\infty}$ and choosing the potential reference there to $0$, it reads 
\begin{align}\label{eqn:app_density_ele}
      n_{\rm{e}}=n_{\infty} \exp\left(\frac{e\Phi}{T_{\rm{e}}}\right).
\end{align}
The ion flux to the probe is spatially constant. We write is as a product of the density $n_\infty$, the Bohm velocity $\sqrt{T_{\rm{e}}/m_{\rm{i}}}$, and an
yet unknown constant $\hat{R}$:
\begin{align}
     4\pi r^2n_{\rm{i}}v_{\rm{i}}= -  4\pi\hat{R}^2 n_{\infty} \sqrt{\frac{T_{\rm{e}}}{m_{\rm{i}}}}.
\end{align}
The ion velocity can be derived from the energy conservation
\begin{align}
      \frac{1}{2}m_{\rm{i}} v_{\rm{i}}^2 + e \Phi = 0.
\end{align}
The ion density is then obtained as
\begin{align}\label{eqn:app_density_ion}
      n_{\rm{i}}=\frac{\hat{R}^2}{r^2}
      \,n_{\infty}\sqrt{-\frac{T_{\rm{e}}}{2e\Phi}}.
\end{align}
Inserting \ref{eqn:app_density_ele} and \ref{eqn:app_density_ion} into \ref{eqn:app_Poisson}, we have
\begin{align}\label{eqn:app_diff_eqn}
      - \varepsilon_0\frac{1}{r^2}\frac{\partial}{\partial r} r^2     \frac{\partial\Phi}{\partial r}  =  e n_{\infty} \left(\cfrac{\hat{R}^2}{r^2}\sqrt{-\frac{T_{\rm{e}}}{2e\Phi}}-\exp\left(\frac{e\Phi}{T_{\rm{e}}}\right) \right).
\end{align}
The floating condition states that the electron flux and the ion flux to the probe are equal.
Employing the Hertz-Langmuir  formula~\cite{Riemann1989}, it is demanded that
\begin{align}
    4\pi\hat{R}^2 n_{\infty} \sqrt{\frac{T_{\rm{e}}}{m_{\rm{i}}}}
    = 4\pi R^2
      \sqrt{\frac{T_{\rm{e}}}{2\pi m_{\rm{e}}}}\, n_{\infty} \exp\left(\frac{e\Phi(R)}{T_{\rm{e}}}\right).
\end{align}
For any set of parameters, the 
 unknown $\hat{R}$ can be found by means of the shooting method. Fig~\ref{fig:Density} shows the potential and the density profiles of the electrons and ions for an example. In the simulation, the particles are generated according to the density profiles as the initial condition. 

\end{appendix}

\newpage
\section{Figures}

\newpage

\newpage

\newpage

\newpage

\newpage

\newpage

\newpage

\newpage

\newpage
\end{document}